\documentstyle[preprint,revtex]{aps}

\begin{document}
\draft
\rightline{NSF-ITP-93-81}
\begin{title}
Influence of Gap Extrema on the Tunneling Conductance

Near an Impurity in an Anisotropic Superconductor
\end{title}

\author{J.M. Byers}

\begin{instit}
Department of Physics

University of California, Santa Barbara, CA 93106-9530
\end{instit}

\moreauthors{M.E. Flatt\'e}

\begin{instit}
Institute for Theoretical Physics

University of California,
Santa Barbara, CA 93106-4030
\end{instit}

\moreauthors{D.J. Scalapino}
\begin{instit}
Department of Physics

University of California, Santa Barbara, CA 93106-9530
\end{instit}

\begin{abstract}
We examine the effect of an impurity on the nearby tunneling conductance
in an anisotropically-gapped
 superconductor. The variation of the conductance
has pronounced spatial dependence
which depends strongly on the Fermi surface location of gap extrema.
In particular, different gap
symmetries produce profoundly different spatial features in the conductance.
These effects may be detectable with an STM measurement on the surface of
a high-temperature superconductor.
\end{abstract}

\narrowtext

Any well-developed theory of high-temperature superconductivity must
predict the symmetry of the energy gap. Some proposals, including
the strongly-coupled phonon-mediated pairing which
causes superconductivity in ordinary metals\cite{Eliashberg}, yield
isotropic or
nearly isotropic gaps. The dominant characteristic of the pairing
interaction in this model is retardation. In the low-temperature
superconductors this is reflected in the frequency-dependence of the
superconducting gap and clearly shows that phonons mediate the pairing.

For a nearly half-filled Hubbard model,
approximations involving the exchange of para-antiferromagnetic spin
fluctuations give rise to an effective electron-electron interaction
which has a characteristic momentum dependence, peaking near the
antiferromagnetic wave vector\cite{Varma,Phillippe}.
A similar momentum dependence has been found
in Monte Carlo simulations\cite{MC}. Such an interaction favors
$d_{x^2-y^2}$ pairing and gives rise to a momentum-dependent gap
which has four nodes.

Gap-measurement techniques that were introduced and developed for use
on electron-phonon superconductors measure accurately the frequency
structure of the gap.
Among these are the voltage dependence of the tunneling $I(V)$
characteristic into the
homogeneous superconductor\cite{tunnel} and the frequency-dependence of
electromagnetic absorption\cite{ref}.
If, however, the gap has
strong momentum-dependence these measurements depend on some
momentum-averaged
value of the gap. Most of these probes\cite{tunneling,reflect}
now indicate
pronounced gap anisotropy or gapless superconductivity in the
high-temperature superconductors.

In the
cuprate-oxide superconductors it would be useful to have experimental
information on the
momentum-dependence of the gap.
Certain techniques exist which are designed to
detect gap nodes on the Fermi surface.
These include measurements of the
low-temperature dependence of thermodynamic and transport
properties, such as the specific
heat\cite{tsh} and the magnetic penetration depth\cite{Dexp}.
Here power-law dependencies on temperature
typically imply nodes, in contrast to the exponential behavior
associated with a fully gapped Fermi surface.
Measurements of these quantities proved extremely
illuminating in studies of
heavy-fermion materials\cite{Broholm}.  Recent
results\cite{Dexp} on
$\rm YBa_2Cu_3O_7$ indicate a low-temperature penetration
depth proportional to $T$, suggesting the existence of nodes.

At this time, however, only a few experiments exist that directly
measure the gap
as a function of momentum on the Fermi surface. Angle-resolved
photoemission (ARPES) on $\rm Bi_2Sr_2CaCu_2O_8$
measures the magnitude of the gap\cite{Shen} with an
angular resolution of about $6^o$ and an energy resolution of about
$10$~meV.  A gap has been detected with ARPES only in $\rm
Bi_2Sr_2CaCu_2O_8$.
Measurements of $\rm YBa_2Cu_3O_7-Pb$
SQUIDs may eventually determine the symmetry
of the superconducting gap\cite{Wollman}, but do not provide detailed
information on the momentum-dependence of the gap.
In theory, measurements
of linewidth changes of phonons with finite momentum have excellent
angular and energy resolution\cite{Flatte}, and can measure the
relative phase of the gap\cite{FlatSQ}. Such measurements have yet to
be done.

Here we propose that a scanning tunneling microscope (STM) study of the
spatial variation of the tunneling conductance \cite{Sauls}
around an impurity
can be used to probe the momentum-dependence of the superconducting gap
$\Delta_{\bf k}$.  We will discuss in this Letter those features of the
gap which produce the most
profound changes in the local tunneling conductance: gap minima and
maxima.
A gap minimum or maximum on the Fermi surface produces strikingly large
conductance variations along the spatial directions away from the
impurity which are perpendicular to the Fermi surface tangent at the gap
extremum.
We numerically
calculate the spatial structure for superconductors with (a) an isotropic
gap, (b) a $d_{x^2-y^2}$ gap, and
(c) an anisotropic but nodeless gap.
The results are explained by analytic forms in
various limits.
The energies of gap maxima and minima should be discernible
with a resolution better than $1$~meV.

The feasibility of observing the local tunneling density of states with
atomic resolution
has been established
by STM tunneling conductance measurements near impurities and step edges
on the surface of Cu(111)\cite{Crommie}. In addition,
tunneling conductances have been obtained with
near-atomic resolution
on the surface of the high-temperature superconductor
$\rm Bi_2Sr_2CaCu_2O_8$\cite{Wolf}.

We assume the impurity has three possible effects on the local
environment in a superconductor: an impurity potential
$\delta\epsilon$, a change in the
superconducting gap $\delta\Delta_{\bf k}$ (which depends on the
momentum {\bf k}), and a change in the
local mass $\delta m$ of the carriers.
In this Letter each of these is modeled as a delta-function perturbation
at an impurity site {\bf 0}, perturbing the
homogeneous Hamiltonian of the bulk superconductor.
An impurity would actually
have a finite-range effect on $\delta\epsilon$, $\delta\Delta_{\bf
k}$, and
$\delta m$,
which could be as large as a superconducting coherence length.
Here we will be concerned with
conductance changes which persist at greater distances.

Each of the
perturbations contribute to the others; for example, introducing an
impurity potential will have an effect of the same order on
the gap magnitude. This in turn affects the conductance. The
precise relationship between the parameters depends upon details of the
microscopic mechanism.
We will use
$\delta\epsilon$ and $\delta\Delta_{\bf k}$ as phenomenological
parameters in our model, and assume that they
satisfy the self-consistent equations for the superconductor.
The effect of a $\delta m$ can be absorbed into
$\delta\epsilon$.
We thus avoid the difficulty which results from
attempting to start from
``bare'' impurity parameters.

The effect of the impurity perturbations on the conductance
under the STM tip can be calculated from the Green's functions of the
homogeneous superconductor.
These are written compactly in the Nambu formalism\cite{Schrieffer}:
\begin{equation}
{\bf G_o}({\bf x},{\bf x'},\omega) = \left(\matrix{G_o({\bf x},
{\bf x'},\omega)&F_o({\bf x}, {\bf x'
},\omega)\cr F_o({\bf x},{\bf x'},
\omega)&-G_o({\bf x},{\bf x'},-\omega)\cr}\right),
\label{Nambu}
\end{equation}
where
\begin{eqnarray}
G_o({\bf x},{\bf x'},
\omega) =&& {1\over {\cal V}
} \sum_{\bf k} {\rm e}^{i{\bf k}\cdot({\bf x}-{\bf
x'})}{\omega + \epsilon_{\bf k}\over \omega^2-\epsilon_{\bf k}^2 -
\Delta_{\bf k}^2 + i\eta}\nonumber\\
F_o({\bf x},{\bf x'},
\omega) =&& {1\over {\cal V}} \sum_{\bf k} {\rm e}^{i{\bf k}\cdot({\bf x}-{\bf
x'})}{\Delta_{\bf k}\over \omega^2-\epsilon_{\bf k}^2 -
\Delta_{\bf k}^2 + i\eta}.
\end{eqnarray}
Here $\epsilon_{\bf k}$ is the single-particle energy
in the normal material, $\Delta_{\bf k}$ is the superconducting gap,
{\bf x} and ${\bf x'}$ are positions and ${\cal V}$ is the volume.
The conductance at a position {\bf r} and voltage $V$
is the imaginary part of the upper-left
entry in the matrix (\ref{Nambu}):
$dI({\bf r},V)
/dV =- A{\rm Im}G_o({\bf r},{\bf r},\omega=V)$. Here $A$ consists of
voltage and position-independent matrix elements and numerical factors.

The Green's function with an impurity at {\bf 0} to linear
order in the perturbations is
\begin{equation}
{\bf G}({\bf r},{\bf r},\omega) = {\bf G_o}({\bf r},{\bf r},\omega) +
{\bf G_o}({\bf r},{\bf 0},\omega)
\left[ \delta\epsilon \tau_3
{\bf G_o}({\bf 0},{\bf r},\omega) + \delta\Delta \tau_1
{\bf G_o'}({\bf 0},{\bf r},\omega)\right],
\end{equation}
where $\tau_1$ and $\tau_3$ are Pauli matrices, and
\begin{eqnarray}
G_o'({\bf x},{\bf x'},\omega)
=&& {1\over {\cal V}} \sum_{\bf k} {\rm e}^{i{\bf k}\cdot({\bf x}-{\bf
x'})}d({\bf k})
{\omega + \epsilon_{\bf k}\over \omega^2-\epsilon_{\bf k}^2 -
\Delta_{\bf k}^2 + i\eta}
\nonumber\\
F_o'({\bf x},{\bf x'},\omega) =&&
{1\over {\cal V}} \sum_{\bf k} {\rm e}^{i{\bf k}\cdot({\bf x}-{\bf
x'})}d({\bf k})
{\Delta_{\bf k}\over \omega^2-\epsilon_{\bf k}^2 -
\Delta_{\bf k}^2 + i\eta}.
\end{eqnarray}
Here $\delta\Delta_{\bf k} = \delta\Delta d({\bf k})$, where the maximum
value of $d({\bf k})$ on the Fermi surface is $1$.
Since the Green's functions in Eq. (2) and Eq. (4) only depend on $|{\bf
x}-{\bf x'}|$, the two position variables {\bf x} and
${\bf x'}$ will be replaced by $|{\bf x} - {\bf x'}|$
below.

It is useful to express the conductance in terms of dimensionless quantities.
We define Green's functions normalized by the
density of states at the Fermi energy $N^*$, such as
$g_o = G_o/N^*$.
Then, in terms of the normalized couplings
$\delta\tilde\epsilon = \delta\epsilon
N^*$ and $\delta\tilde\Delta = \delta\Delta N^*$, the
conductance at {\bf r} is
\begin{eqnarray}
{dI({\bf r},V)\over dV} = -\Bigg[{\rm Im} g_o({\bf 0},V) &&+
\delta\tilde\epsilon
{\rm Im}\left\{g_o^2({\bf r},V) + f_o^2({\bf r},V)\right\} \nonumber\\
&&+ \delta\tilde\Delta
{\rm Im}\left\{g_o'({\bf r},V)f_o({\bf r},V)  + f_o'({\bf r},V)
g_o({\bf r},V)
\right\} \Bigg],\label{conduct}
\end{eqnarray}
where $dI({\bf r},V)/dV$ is normalized by the normal metal's conductance.

We have plotted in
Fig. 1 the normalized conductance (not including the bulk term)
from Eq. (\ref{conduct}) for $\delta\tilde\epsilon=1$ and
$\delta\tilde\Delta = 0$
for two gaps: (a) an isotropic gap,
$\Delta(\phi) = \Delta_o$, (b) a $d_{x^2-y^2}$ gap, $\Delta(\phi)
= \Delta_o\cos 2\phi$.
The bias $V$ is set to $1.1\Delta_o$.
Strong angular-dependence of the signal in Fig. 1(b) is apparent.
We have not plotted the conductance for the
anisotropic s-wave gap
$\Delta(\phi)  = \Delta_o(0.55 + 0.45\cos 4\phi)$, since
at this bias voltage
it appears essentially the same as Fig. 1(b). We will show later that it
can be distinguished from the $d_{x^2-y^2}$ gap
at lower bias voltages, near the anisotropic s-wave
gap minimum. The Fermi surface is
assumed circular, although the gap is considered pinned to the lattice.
Our use of a cylindrical Fermi surface
is not a significant assumption since, as will be
explained below, the main
contribution to any signal comes from small regions of the Fermi
surface.

In order to illustrate the strong angular-dependence of the conductance
for the $d_{x^2-y^2}$ gap,
we show in Figs. 2(a) and (b) the conductance as a function of
distance for two angles. The first angle, $\phi = 0$, is the direction
of a gap maximum. The second angle, $\phi = \pi/4$, is the direction of
a node.  For
comparison we show in Fig. 2(c) the conductance as a function of distance
for an isotropic gap.

We will now analyze the Green's function $g_o$ to understand the
pronounced spatial variation in Figs. 1 and 2.
It is straightforward to reduce the integrations for the Green's
functions to angular integrals. Defining ${\bf r}  = (x,y) = k_F^{-1}
\rho(\cos\phi_o,\sin\phi_o)$,
\begin{eqnarray}
g_o(\rho,\phi_o,&&\omega) = -{i\over 2}\int
d\phi \Bigg[
\left( {\omega \over
[\omega^2 - \Delta^2(\phi)]^{1/2}}
+{\rm sgn}\{\cos(\phi-\phi_o)\}
\right)\Theta(\omega-\Delta(\phi))
\times\nonumber\\
&& \exp\left\{i\rho\left(1+2{\rm sgn}\{\cos(\phi
- \phi_o)\}{[\omega^2
-\Delta^2(\phi)]^{1/2}\over v_Fk_F}\right)^{1/2}\cos(\phi-\phi_o)\right\} +
\nonumber\\ &&
-i\left( {\omega \over
[\Delta^2(\phi) - \omega^2]^{1/2}}
+i{\rm sgn}\{\cos(\phi-\phi_o)\}
\right)
\Theta(\Delta(\phi) - \omega)
\times\nonumber\\
&& \exp\left\{i\rho\left(1+2i{\rm sgn}\{\cos(\phi
- \phi_o)\}{[ \Delta^2(\phi) - \omega^2
]^{1/2}\over v_Fk_F}\right)^{1/2}\cos(\phi-\phi_o)\right\}
\Bigg].
\end{eqnarray}
Here $\Theta(E)$ is the Heavyside step function.
$f_o(\rho,\phi,\omega)$ is given by a similar expression, with the
parenthetical factors before the step functions replaced by
$\Delta(\phi)/|\omega^2-\Delta^2(\phi)|^{{1\over 2}}$.
Similarly, $g_o'$ and $f_o'$ have the same
form as $g_o$ and $f_o$ with an additional factor of $d(\phi)$ in the
integrand.

The largest contributions to the type of integral in Eq. (6) come from
two occurrences in the integrand. The first occurrence is
when the bias $\omega$ is very close to the gap $\Delta(\phi)$. In
this instance the denominator of the integrand becomes small. The
second occurrence is
when the curvature of the exponentiated factor becomes
small. Since $\int d\phi \exp(\alpha\phi^2)\propto \alpha^{-1/2}$, when
the curvature $\alpha$ is small the integral becomes large. To
locate angles where the exponent has small curvature, we have expanded the
$\phi$-dependent factors of the exponent.
The curvature is a
minimum when $\Delta(\phi)$ is a minimum or maximum,
{\it and} $\omega \sim \Delta(\phi)$.
The integration near these angles provides most of the variation in the
tunneling conductance and we find the following form for the local
contribution to the conductance in a $d_{x^2-y^2}$ superconductor:
\begin{eqnarray}
{dI(\rho,\phi_o,V)
\over dV}&& = {-\pi\over \rho\cos\phi_o\alpha}
\left({V^2\over V^2-\Delta_o^2}\right)
\Bigg[ \sin\left\{ 2\rho\cos\phi_o
\left(1+{(V^2-\Delta_o^2)^{{1\over 2}}\over
v_Fk_F}\right)\right\}\times
\label{final}\\
&&\qquad\qquad - {\alpha\over 1-\alpha}\sin\left\{ 2\rho\cos\phi_o
\left(1-{(V^2-\Delta_o^2)^{{1\over 2}}\over
v_Fk_F}\right)\right\}\nonumber\\
&&+ {2\alpha^{{1\over 2}}\over (1-\alpha)^{{1\over 2}}}
\sin\left\{ 2\rho\cos\phi_o
{(V^2-\Delta_o^2)^{{1\over 2}}\over
v_Fk_F}\right\}\Bigg]
{\sin^2(\rho\sin\phi_o\Delta\phi)\over (\rho\sin\phi_o)^2}.\nonumber
\end{eqnarray}
For brevity, the solution for $-\pi/4 < \phi_o< \pi/4$ is presented
above. All other angles $\phi_o$ can be mapped into this region.
$\Delta\phi$ is a $\rho$-independent angle and $\alpha$ is the
curvature at the extrema, which is approximately given by
\begin{equation}
\alpha = {1\over 2} - {2\Delta_o^2\over v_Fk_F
(V^2-\Delta_o^2)^{{1\over 2}}}.
\end{equation}
The conductance in Eq. (\ref{final})
has oscillations whose wavelength depends on the bias. This
bias dependence should help screen out spurious effects.

The isotropic-gap superconductor has no Fermi surface location
where the curvature becomes smaller than elsewhere. Thus we find, as
shown in Fig. 2, that the amplitude of the conductance variations is
larger for the $d_{x^2-y^2}$ gap, in the spatial direction corresponding
to a gap extremum, than the amplitude for the isotropic gap.
The conductance for the isotropic-gap superconductor is given by the
expression in Eq.  (\ref{final}) with the $\sin^2(\rho\sin\phi_o
\Delta\phi)/(\rho\sin\phi_o)^2$ and
$\cos\phi_o$ factors absent and $\alpha=1/2$.

Since the
main contribution to the high-signal regions comes from near the gap
extrema (where the curvature is least), it is reasonable to set $\omega
= \Delta_o = \Delta(\phi)$ in these integration regions. Then
$f_o=g_o=f_o' = g_o'$. The analytic work done on $g_o$ above thus
applies to all the Green's functions. We find numerically
that plots of the change in the tunneling conductance
due to a change in $\delta\tilde
\Delta_{\bf k}$ are almost identical to those for
a change in $\delta\tilde\epsilon$.

In Fig. 3 we show the normalized conductance for
the anisotropic s-wave gap and the $d_{x^2-y^2}$ gap at a
bias $V = .16\Delta_o$ (this is just above the minimum gap for the
anisotropic s-wave gap).
The isotropic gap is not shown because there is
no variation in its conductance due to the impurity at this bias.
Figs. 3(a) and (b)
show the change in the tunneling conductance for the
anisotropic s-wave gap as a function
of distance in the direction of the gap minimum and maximum
respectively. The bias is close to the minimum, so there
is a long-range enhanced signal only for (a). In contrast to (a) and (b),
the change in
conductance for the $d_{x^2-y^2}$ gap in the same directions, shown in
(c) and (d), is small.

Here we have shown that
there is a characteristic structure in the local conductance at bias
voltages near gap minima or maxima which can be used to probe the
momentum-dependence of $\Delta_{\bf k}$. Thus one can obtain information
about the symmetry of the gap. Beyond this, we believe that scanning
tunneling microscopy of the local density of states around an impurity
site can provide detailed information on both the momentum and frequency
dependence of the gap. This would provide a means of determining both the
momentum and frequency dependence of the interaction responsible for
pairing in the cuprate superconductors.

J.M.B. and D.J.S. acknowledge the support of the National Science
Foundation grant DMR90-02492.
M.E.F. acknowledges the support of this work
by the National Science Foundation under Grant No. PHY89-04035.

\figure{ Normalized conductance for a unit perturbation to the site energy
$\delta\tilde\epsilon$. The length scale is set by
$k_F^{-1}$.
Here we have taken a
coherence length equal to
$10k_F^{-1}$. The voltage is $1.1\Delta_o$, where $\Delta_o$ is
the gap maximum. The range of the intensity is from -0.3 (black) to 0.3
(white).  (a) isotropic gap, $\Delta(\phi) = \Delta_o$
(b) $d_{x^2-y^2}$ gap, $\Delta(\phi) = \Delta_o\cos 2\phi$.}

\figure{ Same function as Fig. 1. (a) Radial plot of the change in
the conductance due to $\delta\tilde\epsilon$ along the
high-signal direction $\phi=0$ for the $d_{x^2-y^2}$ gap. (b) Radial
plot along the low-signal direction, $\phi = \pi/4$. (c) Radial plot for
the isotropic gap. The scale is the same for all three plots.}

\figure{ Normalized conductance for a unit perturbation to the site
energy $\delta\tilde\epsilon$. Here the voltage, $0.16\Delta_o$,
is just above the gap minimum for the case of an anisotropic s-wave gap
$\Delta(\phi) = \Delta_o(0.55 +0.45\cos 4\phi)$. Radial plots of the
change in conductance due to $\delta\tilde\epsilon$ for the anisotropic
s-wave gap are shown in the direction of the gap minimum (a) and gap
maximum (b). Since the bias is close to the minimum, only (a) has a
long-range enhanced signal. For comparison, the same functions for the
$d_{x^2-y^2}$ gap are shown in (c) and (d). Compared to (a), all three
of (b), (c) and (d) are small. Thus we find a strong dependence of the
tunneling conductance on gap minima, similar to the dependence on gap
maxima shown in Figs. 1 and 2.}


\begin{references}

\bibitem{Eliashberg} G.M. \'Eliashberg, {\it J. Exptl. Theoret. Phys.
(USSR)} {\bf 38}, 966 (1960), translated in {\it Soviet Phys. JETP}
{\bf 11}, 696 (1960).

\bibitem{Varma} K. Miyake, S. Schmitt-Rink, C.M. Varma, {\it Phys. Rev.
B} {\bf 34}, 6554
(1986). D.J. Scalapino, E. Loh, Jr., J.E. Hirsch, {\it Phys. Rev. B}
{\bf 34}, 8190 (1986).

\bibitem{Phillippe}  N.E. Bickers, D.J. Scalapino, R.T. Scalettar, {\it
International Journal of Modern Physics B} {\bf 1}, 687 (1987); N.E. Bickers,
D.J. Scalapino, S.R. White, {\it Phys. Rev. Lett.} {\bf 62}, 961 (1989).
More recently, P. Monthoux, A.V. Balatsky, D. Pines, {\it Phys. Rev.
Lett.} {\bf 67}, 3448
(1991); P. Monthoux, D. Pines, {\it Phys. Rev. Lett.} {\bf 69}, 961 (1992).

\bibitem{MC}  N. Bulut, D.J. Scalapino, S.R. White, {\it Phys. Rev. B}
{\bf 47}, 6157 (1993).

\bibitem{tunnel} W.L. McMillan, J.M. Rowell, in {\it Superconductivity I},
edited by R.D. Parks (Marcel Dekker, New York, 1969), p. 561.

\bibitem{ref} R.R. Joyce, P.L. Richards, {\it Phys. Rev. Lett.} {\bf
24}, 1007 (1970).

\bibitem{tunneling} e.g. D. Mandrus, {\it et al.}, {\it Europhys. Lett.}
{\bf 22}, 199 (1993).

\bibitem{reflect} e.g. M.J. Sumner, J.-T. Kim, T.R. Lemberger, {\it Phys.
Rev. B} {\bf 47}, 12248 (1993). D. Mandrus, {\it et al.}, {\it Phys.
Rev. Lett.} {\bf 70}, 2629 (1993).

\bibitem{tsh} S.E. Stupp, W.C. Lee, J. Giapintzakis, D.M. Ginsberg, {\it
Phys. Rev. B} {\bf 45}, 3093 (1992).

\bibitem{Dexp} W.N. Hardy, D.A. Bonn, D.C. Morgan, R. Liang, K. Zhang,
preprint.

\bibitem{Broholm} C. Broholm, {\it et al.}, {\it Phys. Rev. Lett.}
{\bf 65}, 2062 (1990).

\bibitem{Shen} B.O. Wells, {\it et al.},
{\it Phys. Rev. B} {\bf 46}, 11830 (1992).
Z.-X. Shen, {\it et al.}, {\it Phys. Rev. Lett.} {\bf 70}, 1553 (1993).

\bibitem{Wollman} D.A. Wollman, {\it et al.}, preprint.

\bibitem{Flatte} M.E. Flatt\'e, {\it Phys. Rev. Lett.} {\bf 70}, 658 (1993).

\bibitem{FlatSQ} M.E. Flatt\'e, S. Quinlan, D.J. Scalapino, NSF-ITP-93-25.

\bibitem{Sauls} C.H. Choi, P. Muzikar, {\it Phys. Rev. B} {\bf
41}, 1812 (1990) and T.A. Tokuyasu, D.W. Hess, J.A. Sauls, {\it Phys. Rev. B}
{\bf 41}, 8891 (1990) have studied the anisotropic spatial variations
in the superfluid density and current around an impurity and vortex
respectively in an anisotropic superconductor.

\bibitem{Crommie} M.F. Crommie, C.P. Lutz, D.M. Eigler, {\it Nature}
{\bf 363}, 524 (1993).

\bibitem{Wolf} A. Chang, {\it et al.}, {\it Phys. Rev. B} {\bf 46},
5692 (1992).

\bibitem{Schrieffer} e.g. J.R. Schrieffer, {\it Theory of
Superconductivity} (Benjamin/Cummings, Reading Mass., 1964).

\end{references}
\end{document}